\documentstyle[preprint,aps,exscale,floats,amsfonts,amssymb,graphicx]{revtex}

\newcommand{\D}{\text{d}} 
\newtheorem{lemma}{Lemma}

\begin{document}

\title{Optimal use of time dependent probability density data to
  extract potential energy surfaces} 
\author{Lukas Kurtz} 
\address{MPI f{\"u}r  Quantenoptik, Hans-Kopfermann Str.~1, 85748
  Garching, Germany}  
\author{Herschel Rabitz\thanks{hrabitz@princeton.edu}} 
\address{Department of Chemistry, Princeton University, Princeton, New 
  Jersey 08544-1009, USA}
\author{Regina de Vivie-Riedle\thanks{rdv@mpq.mpg.de}}
\address{MPI f{\"u}r Quantenoptik, Hans-Kopfermann Str.~1,  85748
  Garching, Germany}  

\date{November 09, 2001} 
\maketitle

\begin{abstract}
  A novel algorithm was recently presented to utilize emerging time
  dependent probability density data to extract molecular potential
  energy surfaces. This paper builds on the previous work and seeks to
  enhance the capabilities of the extraction algorithm: An improved
  method of removing the generally ill-posed nature of the inverse
  problem is introduced via an extended Tikhonov regularization and
  methods for choosing the optimal regularization parameters are
  discussed. Several ways to incorporate multiple data sets are
  investigated, including the means to optimally combine data from
  many experiments exploring different portions of the potential. In
  addition, results are presented on the stability of the inversion
  procedure, including the optimal combination scheme, under the
  influence of data noise. The method is applied to the simulated
  inversion of a double well system to illustrate the various points.
\end{abstract}

\section{Introduction}\label{sec:intro}
To fully understand chemical dynamics phenomena it is necessary to
know the underlying potential energy surfaces
(PES)~\cite{LevineBernstein}. Surfaces can be obtained by two means:
{\it ab initio}
calculations~\cite{BornOppenheimer,Hartree,Fock1,Fock2,SzaboOstlund}
and the inversion of suitable laboratory
data~\cite{Morse_PhysRev1929,Gerber_PhysRevLett1978,Lowe_SIAM1992,Ho_JPhysChem1993,GerambInversionTheory,Fabiano_IMA1995,Zhang_JChemPhys1995,Ho_JChemPhys1996}.
This paper is concerned with an emerging class of laboratory
data~\cite{Zewail_JPhysChem1993,Shapiro_JPhysChem1996,Geiser_JChemPhys1998}
with special features for inversion purposes.  Traditional sources of
laboratory data for inversion produce an indirect route to the
potential requiring the solution of Schr{\"o}dinger's
equation~\cite{SchroedingerGleichung} in the process.  An alternative
suggestion~\cite{Zhu_JChemPhys1999Jul,Zhu_JPhysChemA1999} has been put
forth to utilize ultrafast probability density data from diffraction
observations or other
means~\cite{Vager_Science1989,Kwon_PhysRevLett1996,Assion_PhysRevA1996,Williamson_Nature1997,Krause_PhysRevLett1997,Jones_PhysRevA1998}
to extract adiabatic potential surfaces.  Such data consists of the
absolute square of the wavefunction. Although the phase of the overall
wavefunction is not available, there is sufficient information in this
data to extract the potential fully quantum mechanically {\em without}
the solution of Schr{\"o}dinger's equation.  Instead, the proposed
procedure rigorously reformulates the inversion algorithm as a linear
integral equation utilizing Ehrenfest's
theorem~\cite{EhrenfestTheorem} for the position operator.  Additional
attractive features of this algorithm are (a) the procedure may be
operated non-iteratively, (b) no knowledge is required of the
molecular excitation process leading to the data and (c) the regions
where the potential may be reliably extracted are automatically
revealed by the data.

Extensive efforts are under way to achieve the necessary temporal and
spatial resolution of the probability density data necessary for
inversion processes as well as for other
applications~\cite{Zhu_JPhysChemA1999}.  In anticipation of these
developments a number of algorithmic challenges require attention to
provide the means to invert such data. This paper aims to build on the
previous work~\cite{Zhu_JChemPhys1999Jul} and address some of these
needs. In particular this paper will consider (i) optimal choices for
regularizing the inversion procedure, (ii) incorporation of multiple
data sets and (iii) inclusion of data sampled at discrete time
intervals. These concepts are developed and illustrated for the
simulated inversion of a double well potential.

The paper is organized as follows. The basic inversion procedure and
the model system are given in Section~\ref{sec:inversion_scheme}.
Based on the inversion algorithm derived in
Ref.~\cite{Zhu_JChemPhys1999Jul} an extended regularization procedure
is presented in Section~\ref{sec:regularization} followed by a
discussion of a modified time integration scheme applicable to
different types of experimental data sampling. This development
naturally leads to consideration of an optimal combination of data
from different measurements. A proof on how to optimally combine the
data is given in Appendix~\ref{sec:optimality_proof}. The stability of
this data combination procedure under the influence of noise is
discussed as well. Section~\ref{sec:summary} summarizes the findings
of this paper.

\section{The basic inversion procedure and the model
  system}\label{sec:inversion_scheme}

The algorithms developed in this paper will be illustrated for a
one-dimensional system but the generalization to higher dimensions is
straightforward~\cite{Zhu_JPhysChemB2000}: the major difference with
higher dimensions is the additional computational effort involved.
Atomic units are used throughout this work.

For a system whose dynamics is governed by the Schr{\"o}dinger
equation
\begin{equation}
  \label{eq:Schroedinger}
  \text{i}\frac{\partial}{\partial t} \psi(x,t)= 
  \left[-\frac{1}{2m}\frac{\partial^2}{\partial x^2}+V(x)\right]
  \psi(x,t) 
\end{equation}
the time evolution of the average position obeys Ehrenfest's theorem
\begin{equation}
  0= m\frac{\D^2}{\D t^2} \int\! x\, \rho(x,t) \, \D x 
  + \int\! u(x)\,\rho(x,t)\, \D x \;,
  \label{eq:Ehrenfest}
\end{equation}
where $u(x)=\text{d} V(x)/{\text{d} x}$ and
$\rho(x,t)=|\psi(x,t)|^2$. In this work the probability density
$\rho(x,t)$ is assumed to be observed in the laboratory and the goal
is to determine the potential energy surface (PES) $V(x)$ from the
gradient $u(x)$.

Following~\cite{Zhu_JChemPhys1999Jul}, Eq.(\ref{eq:Ehrenfest}) can be
used to construct a Gaussian least squares minimization problem to
determine the PES gradient $u(x)$
\begin{equation}
  J_0\{u(x)\}=\frac{1}{T}\int\limits_{0}^{T}
  \left[\int\! u(x)\,\rho(x,t)\,\D x + m \frac{\D^2}{\D t^2}
    \int\! x\,\rho(x,t) \,\D x \right]^2 \D t\;.
  \label{eq:J0_def}
\end{equation}
The time averaging acts as a filtering process to increase inversion
reliability by gathering together more data.  This will generally
increase reliability which in principle is only limited by the
exploratory ability of the wavepacket.  Beyond some point in time
little information on the potential may be gained by taking further
temporal data starting from any potential initial condition.

Variation with respect to $u(x)$ results in a Fredholm integral
equation of the first kind
\begin{equation}
\frac{\delta J_0\{u(x)\}}{\delta u(x)} =0 \quad\Rightarrow\quad
 \int \! A(x',x)\, u(x')\, \D x' = b(x)
\label{eq:orig_inverseproblem}
\end{equation}
with righthand side (RHS)
\begin{equation}
b(x)=-\frac{m}{T} \int\limits_{0}^{T}\! \rho(x,t)\,
\frac{\D ^2}{\D t^2} \int\! x'\,\rho(x',t) \, \D x' \, \D t
\label{eq:b_def}
\end{equation}
and symmetric, positive semidefinite kernel
\begin{equation}
A(x',x)=\frac{1}{T}\int\limits_{0}^{T}\!\rho(x',t)\rho(x,t)\, \D t\;.
\label{eq:A_def}
\end{equation}
Treated as an inverse problem, Eq.(\ref{eq:orig_inverseproblem})
produces the desired PES gradient $u(x)$ as its solution. For
numerical implementation we resort to the matrix version and its
formal solution
\begin{equation}
\bbox{A} \cdot \bbox{u} \,\Delta x=
\bbox{b}\quad\Rightarrow\quad\bbox{u} = \bbox{A}^{-1}\cdot \bbox{b}
\,\Delta x^{-1};.   
\label{eq:orig_inverseproblem_matrix}
\end{equation}
Here the integral in Eq.(\ref{eq:orig_inverseproblem}) is evaluated at
points of equal spacing $\Delta x$.

This approach to seeking the PES has a number of attractive
features~\cite{Zhu_JChemPhys1999Jul}. The formulation requires no
knowledge of any preparatory steps to produce a specific $\psi(x,0)$
which evolves freely to produce $\rho(x,t)$. The generation of
$A(x,x')$ and $b(x)$ depends only on $\rho(x,t)$ and begins when the
observation process is started.  Moreover, although this is a fully
quantum mechanical treatment there is no need to solve
Schr{\"o}dinger's equation to extract the PES.  The dominant entries
of $A(x,x')$ and $b(x)$ automatically reveal the portions of the PES
that may be reliably extracted. The linear nature of
Eq.(\ref{eq:orig_inverseproblem}) is very attractive from a practical 
perspective.

Notwithstanding these attractions, a principal problem to manage is
the generally singular nature of the kernel of the integral equation
in Eq.(\ref{eq:orig_inverseproblem}).  The kernel's nullspace makes it
difficult to solve the inverse problem and leads to an unstable and
ambiguous solution $u(x)$, two characteristics that generally define
the ill-posedness of inverse problems.  There are two major reasons
for the ill-posedness of the inverse problem in
Eqs.~(\ref{eq:orig_inverseproblem})
and~(\ref{eq:orig_inverseproblem_matrix}).  Firstly, it is not
possible to continuously monitor the wavepacket with arbitrary
accuracy and information is lost due to discrete data sampling in
space and time. Secondly, the ill-posedness is due to the wavepacket
only exploring a subspace of the PES. In regions untouched by the
wavepacket with $\rho(x,t)\!\approx\!0$ for all observation times $t$
the kernel entries vanish as $A(x,x')\! =\! A(x',x)\!=\! \frac{1}{T}
\int_0^T \rho(x,t) \rho(x',t)\,\D t\!\approx\!0$.  Hence these regions
correspond to zero-entry rows and columns in the kernel matrix
$\bbox{A}$ and constitute its nontrivial nullspace.  In general, the
solution $u(x)$ will only be reliable in regions where $\rho(x,t)$ has
significant magnitude during its evolution.  The inversion procedure
can manage the null space with the help of a suitable regularization
procedure.  Singular value decomposition and iterative solution
schemes are available (cf.~\cite{Louis_GAMM1990,NumericalRecipesC} for
an overview), but here we will employ extended Tikhonov regularization
(see Section~\ref{sec:regularization}).  \bigskip

The procedures developed in this paper are applied to a simulated
inversion with a system taken to have a slightly asymmetric double
well potential~\cite{Doslic_JPhysChem1998}
\begin{equation}
  \label{eq:double_well}
  V(x)=\frac{\Delta}{2q_0}(x-q_0) + \frac{\hat{V} -\Delta/2}{q_0^4}
  (x-q_0)^2(x+q_0)^2+\Delta 
\end{equation}
with parameters
\begin{eqnarray}
  q_0    &=& 1.0\\
  \Delta &=& 0.000\,257\quad\text{(asymmetry)} \\
  \hat{V}&=& 0.006\,25\phantom{7} \quad\text{(barrier height)}\;.
\end{eqnarray}
In the work of N.~Do\v{s}li{\'c} {\it et
  al.}~\cite{Doslic_JPhysChem1998} this PES represents a one
dimensional model for the intramolecular proton transfer in
substituted malonaldehyde (see Fig.~\ref{fig:malonaldehyd}). The
particle mass is accordingly that of hydrogen.

The wavepacket propagations to obtain the simulated $\rho(x,t)$ data
employed the split operator method
(cf.~\cite{Feit_JApplOpt1978,Feit_JCompPhys1982}). For propagation as
well as inversion we used a grid with 8192 points over the range
$-4.0\leqslant x\leqslant 4.0$. A time step $\Delta t_{\text{prop}}=3$
was chosen and total propagation time was $T=1200$. The small values
of $\Delta t_{\text{prop}}$ and $\Delta x_{\text{prop}}$ ensured good
convergence of the numerical propagation procedure.

The initial wavefunctions were normalized Gaussian wavepackets of
width $\sigma=0.05$.  As stated earlier, the inversion algorithm
requires no knowledge of how these packets were formed, but generally
one may assume that a suitable external laser field was applied for
times $t<0$. The initial packets were placed at the left (L) and right
minimum (R) of the PES, on top of the barrier (T), and at a location
high on the potential (H).  The wavepacket positions are illustrated
in Fig.~\ref{fig:malonaldehyd} and their exact values, the associated
average energies and the classical turning points at these energies
are given table~\ref{tab:energy}.  The inversion process employed a
time step and grid spacing that differed from those used in the
propagation, as high spatial and temporal resolution is difficult to
attain in the laboratory. Hence, we employed only a portion of all the
available propagation data $\rho(x,t)$ in time and space. We will
present inversion results using every 16th propagation grid point
(i.e., $\Delta x=16\cdot\Delta x_{\text{prop}}$) and every fifth
available snapshot (i.e., $\Delta t =5\cdot\Delta t_{\text{prop}}$);
even fewer snapshots could be used over a longer period of time with
the criterion that roughly the same total amount of data is retained.
The inversion results from these lower resolution data are very
encouraging.

The kernel matrices $\bbox A$ for condition H and T are shown in
Fig.~\ref{fig:kernel}; similar plots apply to the cases L and R. The
kernels are symmetric with respect to $x\!=\!x'$ and their values
cover a large dynamic range from $\thicksim 10^3$ down to $10^{-8}$ on
the plotted domain. Significant entries are found predominantly on the
matrix diagonal, close to the origin of the wavepacket, and also in
the vicinity of the classical turning points.  Beyond the classical
turning points at a distance of approximately $\pm 2.0$ the kernel
values fall off very rapidly for both configurations.

For configuration H in Fig.~\ref{fig:kernel}a the initial narrow
gaussian is peaked at the hydrogen distance $x_0\!=\!1.75$ with
corresponding large entries around $(x,x')\!\approx\!(2,2)$.  The
wavepacket starts to spread and acquires momentum as it slides down
the PES, which results in the broadening diagonal trace observed as
the central structure in Fig.~\ref{fig:kernel}a.  When the wavepacket
reaches its lefthand turning point it spreads further (star structure
around $(x,x')\!\approx\!  (-1.75,-1.75) $) before it returns.  This
pattern coincides with the motion of the average position $\langle
x(t)\rangle$ displayed for configuration H in
Fig.~\ref{fig:individual_reconstruction}a.

Even higher symmetry can be observed for configuration T's kernel
matrix $\bbox{A}$ in Fig.~\ref{fig:kernel}b.  The initial gaussian
remains centered around $x_0\!=\!0.0052$ and spreads to the left and
righthand well only. This is further supported by the motionless
average position $\langle x(t)\rangle$ in
Fig.~\ref{fig:individual_reconstruction}a. Hence large entries in
$\bbox{A}$ result in the vicinity of $(x,x')\!=\!(x_0,x_0)$ and the
wavepacket's symmetrical spread to the left and righthand side of the
PES produces the spikes along the $x$-axis for $x'\!=\!0$.  Due the
kernel's symmetry these spikes reappear as lines along the $x'$-axis
for $x\!=\!0$. Large contributions for $x\!=\!x'$ will again lead to a
pronouced diagonal and add to the snowflake appearance of
Fig.~\ref{fig:kernel}b.

The features of the kernels in Fig.~\ref{fig:kernel} coincide with the
nature of the inverse problem mentioned earlier: symmetry,
ill-posedness, and automatic identification of the range where the PES
may be be reliably extractable (i.e., where the kernel entries are
large).  For configuration H the relevant range is $-2\lesssim
x\lesssim 2$ and for configuration T only the vicinity of the barrier
top should yield reliable PES information. In both cases we cannot
expect reasonable solutions beyond $\pm 2.0$, which coincides with the
classical turning points given in table~\ref{tab:energy}.

\section{An improved regularization procedure}\label{sec:regularization}

Tikhonov regularization~\cite{TikhonovJohn} is straightforward to
implement with simple control provided by suitable weight parameters.
It provides a well defined means to stabilize the
inversion and extract reliable PES information in those regions
allowed by the data.

This investigation goes beyond the initial
work~\cite{Zhu_JChemPhys1999Jul} to carefully explore various
regularization options. Regularization has the goal of improving the
accuracy of the solution, assuring stability and ease of use including
computational simplicity.  The functional $J_0$ was augmented by a
regularization term involving a set of increasingly higher order
differential operators acting on $u(x)$
\begin{equation}
J_1\{u(x)\}=J_0\{u(x)\} + \sum_{\nu = 0}^N \alpha_\nu \xi^{-1}
\int \left[ \left( \xi\frac{\D}{\D x} \right)^\nu u(x) \right]^2 \,\D x\;,
\label{eq:J1_def}
\end{equation}
with real coefficients $\alpha_\nu>0$ and a reference length $\xi$. In
practice $\xi$ may be thought of as the spatial resolution of the data
and in the present numerical simulation it was taken as $\Delta x$.
For a multidimensional system, $\xi$ and $\alpha_\nu$ will become
direction dependent tensors. The parameter $\xi$ acts to ensure that
all the new terms added to $J_0$ have the same units as $[u]^2$ as
well as permits comparison of the roles of the dimensionless
regularization parameters $\alpha_\nu$ for different $\nu$ and
different grid spacings $\Delta x$.

The previous work~\cite{Zhu_JChemPhys1999Jul} did not employ a
reference length as only the $\nu=0$ regularization term was
considered. The parameter $\alpha_0$ penalizes the value of $u(x)$.
The new terms go beyond and impose extra pressure on the gradient
($\nu=1$), the curvature ($\nu=2$) of $u(x)$, etc.\,.

Variation of $J_1$ with respect to $u(x)$ yields the modified
inversion prescription
\begin{equation}
  \int\! \left[ A(x',x) +\delta(x-x')\cdot\sum_{\nu = 0}^N \alpha_\nu
    \xi^{-1} \left( -\xi^2\frac{\D ^2}{\D x'^2}\right)^\nu
  \right]\,u(x')\, \D 
  x' = b(x)\;. 
  \label{eq:new_reg_inverseproblem}
\end{equation}
The sum added to $J_0$ in Eq.(\ref{eq:J1_def}) for regularization
consisted of purely positive terms with derivatives of up to $N$th
order, resulting in an alternating series of only even derivatives up
to order $2N$ in Eq.(\ref{eq:new_reg_inverseproblem}).  Moreover, the
Fredholm integral equation of the first kind has been transformed into
an integro-differential equation for $u(x)$ with the added terms
dominating in the regions where kernel is singular.

Due to the rapid growth in the order of the derivatives it is often
sufficient to set $N=2$, i.e., retaining standard, gradient, and
curvature Tikhonov regularization.  For numerical application
Eq.(\ref{eq:new_reg_inverseproblem}) may be transformed into the
matrix problem
\begin{equation}
\left[\bbox A+\alpha_0\Delta x^{-2}\,\openone 
-\alpha_1\,\bbox D
+\alpha_2(\Delta x)^2\,\bbox Q\right]\cdot\bbox u\,\Delta x= \bbox b\,,
\label{eq:orig_reg_inverseproblem_matrix}
\end{equation}
employing the unit matrix $\openone$, as well as the second
\begin{equation}
\bbox D=\frac{1}{(\Delta x)^2}
\left( \begin{array}{rrrrr}
    -2&1&&&0\\
    1&-2&1&&\\
    &\ddots&\ddots&\ddots&\\
    &&1&-2&1\\
    0&&&1&-2
  \end{array} \right)\;,
\label{eq:2nd_order_differentiation_matrix}
\end{equation}
and the forth order differentiation band matrices
\begin{equation}
\bbox Q=\frac{1}{(\Delta x)^4}
\left( \begin{array}{rrrrrrrrr}
    6&-4&1&0&&&&\cdots&0\\
    -4&6&-4&1&0&&&&\vdots\\
    1&-4&6&-4&1&0\\
    0&1&-4&6&-4&1&0\\
    \vdots&\ddots&\ddots&\ddots&\ddots&\ddots&\ddots&\ddots
  \end{array} \right)\;.
\label{eq:4th_order_differentiation_matrix}
\end{equation}
These are simple differencing expressions for the derivatives
involved. Higher order expressions for the derivatives could be
considered, but finite data resolution and laboratory noise will
generally not warrant or support the added complexity.

To investigate the inverse solution's dependence on the various
regularization parameters in
Eq.(\ref{eq:orig_reg_inverseproblem_matrix}) several parameter scans
for all four configurations L, T, R, H were performed for different
resolutions $\Delta x$ and combinations of $\alpha_\nu$-parameters.
For the discussion in this paper, we selected typical results for the
situation of H with $\Delta x=16\Delta x_{\text{prop}}$. The curves in
Fig.~\ref{fig:scans} show the solution defect $|\Delta u|$ and the
system defect $|\Delta s|$ as defined below in
Eqs.(\ref{eq:solution_defect}) and (\ref{eq:rhs_defect}).  While only
$|\Delta s|$ is an experimentally accessible figure of merit, an
investigation of $|\Delta u|$ here allows for quantifying the quality
of the inverse solution. For both error measures reported the plots
are generated for each $\alpha_\nu$ independently while the others are
kept zero.

Figures \ref{fig:scans}a and \ref{fig:scans}b display the solution defect
\begin{equation}
  \label{eq:solution_defect}
  |\Delta u|=\left[\frac{1}{x_b-x_a}
    \int\limits_{x_a}^{x_b}\left(\,u_{\text{exact}}(x)-u(x)
      \,\right)^2\,\D x\right]^{1/2}\;.
\end{equation}
Figure~\ref{fig:scans}a is computed with $x_a=-2.0$, $x_b=2.0$ (i.e.
the central domain indicated in Fig.~\ref{fig:kernel} and
table~\ref{tab:energy} within which the inversion is expected to be
valid) and Fig.~\ref{fig:scans}b with $x_a=-4.0$, $x_b=4.0$ (i.e., the
full simulation range). The differences between the two cases are
striking.  The corresponding solution defects show a completely
different shape with minima that differ by several orders of magnitude
in $\alpha_\nu$.  In Fig.~\ref{fig:scans}b the magnitude of the error
in the active domain $-2\lesssim x\lesssim 2$ is overestimated. This
behavior in Fig.~\ref{fig:scans}b is due to large deviations between
the exact gradient and the inversion solution for the gradient, which
cannot be recovered reliably in the domain's outer limits. Thus we
conclude that $|\Delta u|$ scans should only be computed over the
regions actually reached to a significant degree by the wavepacket
(cf., Fig.~\ref{fig:scans}a) to achieve reliable estimates of the
inversion quality.

The latter point is illustrated in
Figs.~\ref{fig:individual_reconstruction}b and
\ref{fig:individual_reconstruction}c with the inverted results for
$u(x)$ ane $V(x)$ with pure $\alpha_1$ regularization of
configurations H/H$_1$ where $\alpha_1$ is given in
table~\ref{tab:scan}.  The two cases H/H$_1$ differ in the domain
employed in the inversion (i.e., the active domain for H and the full
domain for H$_1$) and in the choice of optimal $\alpha_1$ determined
according to the $|\Delta u|$ scans.  Thus we further conclude that
the inversion process should be confined to the active domain to
maintain stability.

To find suitable integration regions from the laboratory data the
normalized lefthand
\begin{equation}
  \label{eq:left_variance}
  \sigma^2_{\ell}(t)=
  {\int\limits_{-\infty}^{\langle x \rangle}(x-\langle x
    \rangle)^2\rho(x,t)\,\D x} \left/
    \int\limits_{-\infty}^{\langle x \rangle}2\rho(x,t)\,\D x\right.
\end{equation}
and righthand variance
\begin{equation}
  \label{eq:right_variance}
  \sigma_{\text{r}}^2(t)=
  {\int\limits_{\langle x \rangle}^{\infty}(x-\langle x
    \rangle)^2\rho(x,t)\,\D x} \left/
    \int\limits_{\langle x \rangle}^{\infty}2\rho(x,t)\,\D x\right.
\end{equation}
of the position operator can be helpful. Together with the position
average $\langle x \rangle$ they can provide an estimate for the PES
domain predominantly covered by the wavepacket motion.  We present all
three quantities ($\langle x(t)\rangle$ and
$\sigma_\ell(t),\sigma_{\text{r}}(t)$ as grey shaded regions) in
Fig.~\ref{fig:individual_reconstruction}a. The results clearly show
that for configuration H the range $-2\lesssim x\lesssim2$ is
suitable. For configurations L, T, R an even smaller range is best
(cf., table~\ref{tab:scan}).

All the computations revealed that a gradient Tikhonov regularization
based on $\alpha_1$ performs better than the standard regularization
based on $\alpha_0$ utilized earlier~\cite{Zhu_JChemPhys1999Jul}.
There is some additional improvement in choosing the curvature
regularization $\alpha_2$, but we found it to be less stable for
coarse grids, which will be the standard situation in actual
application.

We also found little improvement in mixing the different
regularization schemes. In general the $\alpha_{\nu}$ regularization
with the largest errors masks the positive effects of the others.
Hence for all cases of the PES reconstruction we utilized only
$\alpha_1$ regularization (cf., the inversion in
Figs.~\ref{fig:individual_reconstruction}a and
\ref{fig:individual_reconstruction}b with the optimal parameters given
in table~\ref{tab:scan}).

As a measure of inversion quality and the role of regularization, we
desire a quantity that is strictly available from the laboratory data
$\rho(x,t)$. A good choice is the system defect $|\Delta s|$ defined
by the norm of satisfying the system
equation~(\ref{eq:orig_inverseproblem}) with the inverse solution
$u(x)$ found via Eq.~(\ref{eq:new_reg_inverseproblem})
\begin{equation}
  \label{eq:rhs_defect}
  |\Delta s|=\left[\frac{1}{L}
    \int\left(\,b(x)- \int A(x,x') u(x')\,\D x' \,\right)^2\,\D
    x\right]^{1/2} \;. 
\end{equation}
The values of $|\Delta s|$ will depend on the regularization
parameters $\alpha_\nu$. Weak regularization will produce a small
value of $|\Delta s|$, but likely artificial structures in the PES.
Over regularization will result in a smooth PES, that is
systematically in error with diminished influence from the kernel
$A(x,x')$ on the inverse solution.  The best choice for the $\alpha_\nu$
is generally where $|\Delta s|$ has risen and leveled off in a stable
region as shown in Fig.~\ref{fig:scans}c. The figure shows that
$|\Delta s|$ naturally tends to zero as $\alpha_\nu\to 0+$ and
monotonically rises until it reaches a plateau. There is very good
agreement between the values of $\alpha_\nu$ which show good results for
$|\Delta u|$ in Fig.~\ref{fig:scans}a and the stable regularization
region identified in Fig.~\ref{fig:scans}c. Thus $|\Delta s|$ should
be of practical utility in assigning regularization parameter values.

The generally self-similar structures in Figs.~\ref{fig:scans}a and
\ref{fig:scans}c suggest that every regularization operator has a
roughly similar effect. This added robustness is also attractive for
practical application if it holds up regardless of the system.

\section{Combining distinct sets of laboratory data}\label{sec:combinations}

Sections~\ref{sec:optimal_rec} and~\ref{sec:rho_combi} will cover
different approaches to combining distinct sets of laboratory density
data. Finally Section~\ref{sec:noise} will explore the impact of data
noise on the inversion.

\subsection{Optimal combination of experimental data}\label{sec:optimal_rec}

The functional $J_0\{u(x)\}$ in its original form in
Eq.(\ref{eq:J0_def}) is expressed in terms of a uniform, continuous
time integration of observed $\rho(x,t)$ data. However, experimental
circumstances including measurements at discrete snapshots in time or
changes in the quality of data sampling may necessitate employing a
weight function $\omega(t)$ for a generalized approach to the time
integration in the functional $J_0$. Thus we define $\hat{J}_0$ as
\begin{equation}
  \label{eq:weighted_J0}
  \hat{J}_0\{u(x)\}=\int\limits_{0}^{\infty}
  \left[\int\! u(x)\,\rho(x,t)\,\D x + m \frac{\D^2}{\D t^2}
    \int\! x\,\rho(x,t) \,\D x \right]^2\omega(t)\, \D t\;.
\end{equation}
The choice $\omega(t) = [\,\Theta(t) - \Theta(t-T)\,]/T$, with
$\Theta$ being the Heaviside step function, will reduce $\hat{J}_0$ to
$J_0$.

Variation of Eq.(\ref{eq:weighted_J0}) leads to a modified inverse
problem
\begin{equation}
  \label{eq:inverse_problem_weighted}
  \frac{ \delta\hat{J}_0\{u(x)\} }{ \delta u(x) }=0
  \quad\Rightarrow\quad
  \int \! \hat{A}(x',x)\, u(x')\, \D x' = \hat{b}(x)\;,
\end{equation}
with the new kernel
\begin{equation}
  \label{eq:kernel_matrix_weighted}
  \hat{A}(x',x)=\int\limits_{0}^{\infty}\!\rho(x',t)\rho(x,t)
 \,\omega(t) \, \D t
\end{equation}
and RHS
\begin{equation}
  \label{eq:rhs_weighted}
  \hat{b}(x)= -m \int\limits_{0}^{T}\!\omega(t)\,
  \rho(x,t)\, \frac{\D ^2}{\D t^2} \int\! x'\,\rho(x',t) \, \D x' \,
  \D t \;.
\end{equation}
The weight $\omega(t)$ does not alter the regularization terms in
Eq.(\ref{eq:new_reg_inverseproblem}). If $\hat{b}(x)$ is rewritten
using partial integration over time, then the weight function must be
considered in this process.

The above equations were applied to two generic cases. First, we
considered data gathered as snapshots in time i.e.,
$\omega(t)=\sum_{j=1}^T\,\delta(t_j-t)$, and evaluated
Eqs.~(\ref{eq:kernel_matrix_weighted}) and~(\ref{eq:rhs_weighted})
with this weight. This procedure simply reduced all time integrations
to sums over the sampled $\rho$ data. Next, we considered the case in
which the measurement process has been divided into two continuous
time intervals of length $T_1$ and $T_2$ separated by a period of time
$\tau$. A reasonable choice of weights would either be
\begin{equation}
  \omega(t)= \frac{\Theta(t)-\Theta(t-T_1)}{T_1+T_2} +
  \frac{\Theta(t-\tau-T_1)-\Theta(t-\tau-T_1-T_2)}{T_1+T_2}\;.
  \label{eq:equal_weighting_intervals}
\end{equation}
or
\begin{equation}
  \label{eq:weighting_intervals}
  \omega(t)= \frac{\Theta(t)-\Theta(t-T_1)}{T_1} +
  \frac{\Theta(t-\tau-T_1)-\Theta(t-\tau-T_1-T_2)}{T_2}\;.
\end{equation}
The choice depends on the desired emphasis to be given to the two data
intervals. Here we chose to give the longer interval a larger
contribution in $\hat{A}(x,x')$ than the shorter one, and this can be
better achieved with using Eq.(\ref{eq:equal_weighting_intervals});
this choice is reasonable, provided the measured data $\rho(x,t)$ in
both intervals are of comparable quality. Clearly many other issues
can be incorporated into the choice of $\omega(t)$ dictated by what is
known about the nature of the data and the information sought about
the PES.

The kernel is now
\begin{equation}
  \label{eq:kernel_intervals}
  \hat{A}(x',x)=\frac{1}{T_1+T_2}
  \left(\int\limits_{0}^{T_1}+\int\limits_{T_1+\tau}^{T_1+\tau+T_2}\right)
\rho(x',t)\rho(x,t)\, \D t
\end{equation}
and the RHS reads
\begin{equation}
  \label{eq:rhs_intervals}
  \hat{b}(x)= -\frac{m}{T_1+T_2} 
  \left(\int\limits_{0}^{T_1}+\int\limits_{T_1+\tau}^{T_1+\tau+T_2}\right)
  \rho(x,t)\,\frac{\D ^2}{\D t^2} \int\! x'\,\rho(x',t) \, \D
    x' \,\D t \;.
\end{equation}

The interpretation of the weight in
Eq.(\ref{eq:equal_weighting_intervals}) is associated with performance
of the inversion with an interrupted gathering of data from a {\em
  single} experiment.  To explore this point further it is useful to
rewrite Eqs.(\ref{eq:kernel_intervals}) and~(\ref{eq:rhs_intervals})
as
\begin{equation}
\int  \left[\,A_1 (x,x')+  A_2 (x,x')\,\right] u(x')\,\D x'
\label{eq:combination}
=b_1(x)+b_2(x)\;,
\end{equation}
where the indices ``1'' and ``2'' denote the evident two data time
domains. In this form the gathering of data from {\em one interrupted}
experiment can also be interpreted as finding the simultaneous
solution to the inverse problem of {\em two different} experiments.
These two experiments could possibly be prepared with distinct
controls could, for example, explore different regions of the PES.

We found that it is optimal to simply combine these sets of data by
addition as indicated in Eq.(\ref{eq:combination}). This procedure
will yield an inverse solution $u_0(x)$ with accuracy greater than a
linear combination $u(x)=\mu u_1(x)+\nu u_2(x)$ of separate solutions
to the individual problems ``1'' and ``2'' as explained below.\bigskip

Consider two experiments that yield two different inverse solutions
satisfying their respective system equation
\begin{equation}
\int A_{1,2} (x,x') u_{1,2}(x')\,\D x' = b_{1,2}(x)\;.
\label{eq:2_experiments_ansatz}
\end{equation}
Naturally there should be only a unique exact $u_{\text{ex.}}(x)$ for
the physical system. Hence both system solutions $u_{1,2}$ in
Eq.(\ref{eq:2_experiments_ansatz}) can be decomposed into the exact
solution and contamination pieces from the kernel's nullspace
\begin{equation}
u_{1,2}(x)=u_{\text{ex.}}(x)+a_{1,2}(x)+r_{1,2}(x)\;.
\label{eq:solution_decomposition}
\end{equation}
The functions $a_{1,2}$ and $r_{1,2}$ are associated with the
nullspace of the two kernels with
$a_{1,2}(x)\in\ker(A_1)\cap\ker(A_2)$ being the contamination from the
common nullspace of $A_1$ and $A_2$ and $r_{1,2}(x)$ the residual
contribution unique to the respective kernel. The goal is to use the
data to find an optimal solution $u_0(x)$ with the smallest possible
nullspace contribution.

Exploiting the linearity of the inverse problem, we may add the two
pieces of Eq.(\ref{eq:2_experiments_ansatz}) to get
\begin{equation}
\int A_{1} (x,x') u_{1}(x')\,\D x' +
\int A_{2} (x,x') u_{2}(x')\,\D x' = b_{1}(x)+ b_2(x)\;.
\label{eq:linearity}
\end{equation}
This doesn't fully satisfy Eq.(\ref{eq:combination}) and it is in
general not possible to construct the optimal solution $u_0(x)$ as a
linear combination $u_0(x)=\mu\,u_1(x) + \nu\,u_2(x)$ with constant
coefficients $\mu, \nu$. To elucidate this point, we insert $u_0(x)$
into Eq.(\ref{eq:combination}) and with the help of
Eqs.(\ref{eq:2_experiments_ansatz})
and~(\ref{eq:solution_decomposition}) we get the cross terms
\begin{eqnarray}
  \int A_{1} (x,x') u_{2}(x')\,\D x' &=& b_{1}(x) + \int A_{1} (x,x')
  r_{2}(x')\,\D x' = b_{1}(x) +{} _1\varepsilon_2(x)\nonumber\\ 
  \int A_{2} (x,x') u_{1}(x')\,\D x' &=& b_{2}(x) + \int A_{2} (x,x')
  r_{1}(x')\,\D x'= b_{2}(x) +{} _2\varepsilon_1(x)\;,
  \label{eq:crossover}
\end{eqnarray}
where the prefactors $\mu$, $\nu$ have been omitted. Hence $u_0(x)$ is
not an optimal solution of Eq.(\ref{eq:combination}) since it leaves
errors $_i\varepsilon_j(x)$ that cannot be eliminated. However, by
employing Eq.~(\ref{eq:combination}) and adding the kernels and RHSs
we can improve the quality of the inversion. No error terms like
$_i\varepsilon_j(x)$ will appear since by construction the resulting
$u_0(x)$ can be decomposed as $u_0(x)=u(x)+a_0(x)$. A contribution
from $r_0(x)$ as in Eq.(\ref{eq:solution_decomposition}) will not
arise, as proved in Appendix~\ref{sec:optimality_proof}.  Thus, the
solution of the combined problem will gain in quality by virtue of the
reduced nullspace of the new kernel $A_1+A_2$.

These optimality results are rigorous but it must be added that in
general any combination of a finite amount of data will not fully
eliminate the nullspace. However in the cases under comparison here
the assumption that a similar degree of robustness can be attained
certainly holds true.

As argued above, we chose the weighting function in
Eq.(\ref{eq:equal_weighting_intervals}) to result in
observation-duration proportional entries in $A_1(x,x')$ and
$A_2(x,x')$. Hence it is quite natural to add $A=A_1+A_2$.  However,
choosing the approach Eq.(\ref{eq:weighting_intervals}) normalizes
each data set independently. This logic naturally leads to considering
the optimal combination of data to form $A=\sigma A_1+\delta A_2$
where $\sigma$ and $\delta$ are positive constants.  This specially
weighted form, or a positive definite combination $A=(1-\beta)
A_1+\beta A_2$ with $\beta\in(0,1)$, might be useful especially in the
presence of different degrees of noise in the two data sets. An
iterative numerical scheme to optimize $\beta$ could then help to
improve the solution by minimizing the effects of nullspace
contamination.\bigskip

The optimal combination of data by addition of kernels $A_i(x,x')$ and
RHSs $b_i(x)$ presented above was applied to the double well system
with results for the gradient $u(x)$ and PES $V(x)$ shown in
Fig.~\ref{fig:combi_reconstruction}. Information was successively
added to the kernel $A(x,x')$ by combining the data sets to form LT,
LTR, and LTRH with the notation based on the initial conditions shown
in Fig.~\ref{fig:malonaldehyd}. In each case all configurations are
weighted equally.  The optimal $\alpha_1$ values employed and defect
measures are given in table~\ref{tab:scan}.

While the individual inverse problem solutions based on L, T, R, and H
reproduce the potential in their respective neighborhoods quite well,
they fail to give adequate results for the other portions of the
potential. On the other hand, the reconstruction of large parts of the
PES is successful if we optimally combine the data of the three
experiments LTR. However, contrary to intuition, we observe that the
solution is less satisfactory from combining all the data LTRH; some
additional oscillations appear along with a dip in the vicinity of the
initial wavepacket for H.  Apparently the nullspace of the expanded
domain cannot be fully managed by $\alpha_1$ regularization alone; no
attempt was made to simultaneously introduce $\alpha_0$ and $\alpha_2$
regularization.

\subsection{Other combinations of data}\label{sec:rho_combi}
Several other schemes for combining the raw density data can be
envisioned, apart from the approach in Section~\ref{sec:optimal_rec}.
One candidate would be the direct combination of $\rho(x,t)$ data from
different experiments. As an illustration we will treat the case of
two different $\rho$'s with
\begin{equation}
  \label{eq:rho_ansatz}
  \rho(x,t)=\rho_1(x,t)+\varepsilon\rho_2(x,t)
\end{equation}
and $\varepsilon$ being a positive constant. This combination is
physically acceptable, as Ehrenfest's theorem in
Eq.(\ref{eq:Ehrenfest}) is linear in the probability density.
Insertion of this sum into the functional $J_0\{u(x)\}$ and variation
with respect to $u(x)$ will yield a formulation analogous to the one
describing inversion under the influence of noise in the data (see
Section~\ref{sec:noise}) in Eq.(\ref{eq:noise_inverse}) upon
comparison of Eqs.(\ref{eq:noise_ansatz}) and (\ref{eq:rho_ansatz}).

The terms proportional to $\varepsilon^0$ and $\varepsilon^2$ will
exactly correspond to what was found earlier in
Eq.(\ref{eq:combination}). However, the terms proportional to
$\varepsilon$ represent a cross correlation between $\rho_1$ and
$\rho_2$. These cross terms can be significant, and they act to
introduce an element of undesirable structure, often oscillatory, in
the equations determining $u(x)$. On physical grounds it is also
artificial to directly correlate the independent experimental data
$\rho_1$ and $\rho_2$ when seeking $u(x)$.

Hence, the scheme of adding together the bare $\rho$-data is expected
to produce unreliable results. To support this argument we present a
test on such a $\rho$-combination consisting of the sum of all four
densities of the initial configurations L, T, R, and H
\begin{equation}
  \rho_\Sigma  (x,t)=\rho_L (x,t)+\rho_T (x,t)+\rho_R (x,t)+\rho_H (x,t)\;.
\end{equation}
The corresponding inverted gradient and PES respectively are shown in
Figs.~\ref{fig:combi_reconstruction}a and
\ref{fig:combi_reconstruction}b. The solution is rather poor and far
worse than the LTRH combination using the same data.  This result
should not be taken to construe that other combinations of data might
not give satisfactory results. However, the combination of $A_i$ and
$b_i$ in Section~\ref{sec:optimal_rec} is quite natural and produces
excellent inversion results.

\subsection{The influence of noise on the inversion}\label{sec:noise}

Any real $\rho$-data will always be contaminated by some degree of
noise. In an additive model this noise contaminated data
$\rho_{\text{n}}(x,t)$ can be represented as
\begin{equation}
  \label{eq:noise_ansatz}
  \rho_{\text{n}}(x,t)=\rho(x,t)+\varepsilon\gamma(x,t)\;,
\end{equation}
where $\varepsilon>0$ is a ordering parameter and the noise is
described by the spatio-temporal function $\gamma(x,t)$.  We assume
that $\gamma(x,t)$ is a randomly varying function with vanishing
average contribution and free from systematic error such that
\begin{equation}
  \label{eq:noise_condition}
  \frac{1}{T}  \int\limits_0^T\gamma(x,t)\sigma(x,t)\,\D t
  \quad\stackrel{T\to\infty}{\longrightarrow}\quad0
\end{equation}
for any function $\sigma(x,t)$ of bounded norm over time that is not
correlated with $\gamma(x,t)$.

Inserting the ansatz in Eq.(\ref{eq:noise_ansatz}) into the functional
$J_0\{u(x)\}$ in Eq.(\ref{eq:J0_def}) and taking the first variation,
the equation determining $u(x)$ is obtained
\begin{eqnarray}
  \label{eq:noise_inverse}
  \lefteqn{\frac{1}{T}\int\!\int\limits_0^T
    \rho(x',t)\rho(x,t) \,\D t\, u(x')\;\D x' 
    +  \frac{\varepsilon}{T}\int\!\int\limits_0^T
    \rho(x',t)\gamma(x,t) \,\D t\; u(x')\,\D x' }
  \nonumber\\
  \lefteqn{  +  \frac{\varepsilon}{T}\int\!\int\limits_0^T
    \gamma(x',t)\rho(x,t) \,\D t\, u(x')\;\D x'  
    +  \frac{\varepsilon^2}{T}\int\!\int\limits_0^T
    \gamma(x',t)\gamma(x,t) \,\D t\; u(x')\,\D x' }
  \nonumber\\
  &=&
  -\frac{m}{T}\int\limits_0^T\rho(x,t)\,
  \frac{\D^2}{\D t^2}  \int x'\rho(x',t)\,\D x'\;\D t
  -\varepsilon\frac{m}{T}\int\limits_0^T\rho(x,t)\,
  \frac{\D^2}{\D t^2}  \int x'\gamma(x',t)\,\D x'\;\D t
  \nonumber\\
  &&  -\varepsilon\frac{m}{T}\int\limits_0^T\gamma(x,t)\,
  \frac{\D^2}{\D t^2}   \int x'\rho(x',t)\,\D x'\;\D t
  -\varepsilon^2\frac{m}{T}\int\limits_0^T\gamma(x,t)\,
  \frac{\D^2}{\D t^2}  \int x'\gamma(x',t)\,\D x'\;\D t\;.
\end{eqnarray}
The terms proportional to $\varepsilon^0$ recover the original
unperturbed system in
Eqs.(\ref{eq:orig_inverseproblem}-\ref{eq:A_def}). Assuming the data
noise level to be small, the terms in $\varepsilon^2$ on both sides of
Eq.(\ref{eq:noise_inverse}) can be neglected.

We first turn to the kernel side of Eq.(\ref{eq:noise_inverse}) and
denote  all terms in $\varepsilon^1$ as the  error kernel $\delta
A(x,x')$
\begin{equation}
  \label{eq:A-error}
  \delta A(x,x')=\frac{\varepsilon}{T}\int\limits_0^T \gamma(x',t)
  \rho(x,t)\, \D t  
  +\frac{\varepsilon}{T}\int\limits_0^T\rho(x',t)\gamma(x,t) \,\D
  t\;.  
\end{equation}
Each term involves the computation of two-point spatial correlations
between functions. However, the functions $\gamma$ and $\rho$ are
uncorrelated, and the temporal integral of their product is expected
to result in only small random contributions to the kernel over $x$
and $x'$, especially for longer time integration as follows from
Eq.~(\ref{eq:noise_condition}).  Following similar logic, the terms
proportional to $\varepsilon^1$ on the RHS of
Eq.(\ref{eq:noise_inverse}) should be negligible, especially for long
time integration. Neglecting the $\varepsilon^2$ terms finally leaves
only the first term proportional to $\varepsilon^0$ on the RHS.

Hence, the functional $J_0$ exhibits some inherent capability to deal
with slightly noisy data. The time integration process averages out
these noise effects so that they should have a decreasing impact on
the inverse solution $u(x)$. Longer periods of temporal data should
make their behavior better.

These results are also in accordance with the stability analysis
presented in~\cite{Zhu_JChemPhys1999Jul}. Resorting to the matrix
version of the inverse problem (cf.,
Eq.(\ref{eq:orig_reg_inverseproblem_matrix})) the authors proved
(Eq.(25) in Ref.~\cite{Zhu_JChemPhys1999Jul}) that the relative error
in the solution ${u}$ after regularization is bounded by the
relative errors in the data $\delta{b}$ and $\delta{A}$.

Moreover it was found (Eqs.(41) and (49)~\cite{Zhu_JChemPhys1999Jul})
that small perturbations in the noise $\varepsilon{\gamma}$ will
result in small proportional perturbations in ${b}$ and ${A}$, which
is excellent behavior for any application with finite time
integration. These results can now be extended to the long time
integration limit where the $\varepsilon^1$ terms in
Eq.(\ref{eq:A-error}) should further diminish in significance for
$T\to\infty$. Similar arguments apply to the
RHS~${b}$~\cite{foot}.

Equation~(\ref{eq:A-error}) also demonstrates why the direct
combination of bare $\rho(x,t)$ data discussed in
Section~\ref{sec:rho_combi} performs less satisfactory than the
optimal combination scheme in Section~\ref{sec:optimal_rec}. In
contrast to the slightly perturbed system cross term $\delta A(x,x')$
above, the analogous term arising from directly combining the $\rho$
data will not vanish. This will introduce an undesirable error
contribution to the inverse problem. In contrast, the optimal
combination scheme for different sets of data in
Section~\ref{sec:optimal_rec} should profit from the inherent
stability of the inversion procedure to deal with slightly noisy
systems since this technique involves a sequence of separate time
integrations.

\section{Summary and outlook}\label{sec:summary}
This paper presented new results that improve and extend a recently
suggested procedure~\cite{Zhu_JChemPhys1999Jul} to extract potential
energy surfaces (PES) from the emerging experimentally observable
probability density $|\psi(x,t)|^2$ data. The results of this paper
should also be applicable to the more general case of extracting the
dipole function from the additional observation of the applied laser
electric field~\cite{Zhu_JPhysChemA1999}.

An easy to implement regularization scheme was introduced, which
increases the accuracy of the computed PES without loss of numerical
stability.  Furthermore an optimal reconstruction method was presented
which combines data from different measurements. This scheme was
argued to be optimal in the sense of reducing the nullspace of the
inverse problem and hence increasing the domain of the extracted PES.
Evidence was presented that this scheme is stable under the influence
of noise, but further investigations will be necessary to fully
confirm these results. We hope that the developments in this paper
stimulate the generation of appropriate probability density data for
inversion implementation.

\acknowledgements

The authors would like to acknowledge Karsten Sundermann who shared
interest in this subject from its inception. RdVR thanks ``Fonds der
Chemischen Industrie'' and HR would like to acknowledge the Department
of Energy. LK acknowledges DFG's financial support through the project
``SPP Femtosekundenspektroskopie''. He also would like to thank
Angelika Hofmann for the propagation code and Jens Schneider as well
as Berthold-Georg Englert for discussions.

\appendix
\section{Optimality proof}\label{sec:optimality_proof}

This section presents the lemma and its proof underlying the optimal
combination of data from different measurements.

\begin{lemma}\label{lemma}
  Given two Hermitian, positive semidefinite operators $\bbox A_1,
  \bbox A_2:{\cal H}\to{\cal H}$ acting on the Hilbert space $\cal H$
  and their sum $\bbox A=\mu\bbox A_1+\nu\bbox A_2$ with coefficients
  $\mu,\nu\in{\Bbb R}>0$, it then holds that
  \begin{displaymath}
  \ker( \bbox A )= \ker(\bbox A_1)\cap\ker(\bbox A_2)\;.
  \end{displaymath}
  For finite dimensional ranges this implies that
  \begin{displaymath}
    {\rm rank}( \bbox A ) ={\rm rank}(\bbox A _1) + {\rm rank}(\bbox A _2)
    -\dim\left(\,{\rm Range}(\bbox A _1)\cap{\rm Range}(\bbox A
      _2)\,\right)\;. 
  \end{displaymath}
\end{lemma}

In other words: Adding two positive semidefinite, Hermitian operators
will reduce the nullspace of the combined operator to that of the
intersection of both nullspaces. The generalization to a finite sum of
operators $\bbox A=\sum_{k=1}^N \alpha_{k}\bbox A_{k}$ with constant
$\alpha_{k}>0$ is evident.\bigskip

Neither positivity nor Hermiticity can be omitted. Without the former
criterion, a counter example is $\bbox A_2=-\bbox A_1$, with
$\mu=\nu=1$. As an example, without the latter criterion, the two
${\Bbb R}^{ 3 \times 3}$ operators
\begin{eqnarray}
  \label{eq:counterexample_hermiticity}
  \bbox A _1=
  \left(\begin{array}{ddd} 1&1&1\\ 0&1&1\\ 0&0&1 \end{array}\right)\;,
  \quad  \bbox A _2=
  \left(\begin{array}{ddd} 0&0&0\\ 1&0&0\\ 1&1&0 \end{array}\right)
\quad\Rightarrow\quad  \bbox A _1+\bbox A_2=
  \left(\begin{array}{ddd} 1&1&1\\ 1&1&1\\ 1&1&1 \end{array}\right)
\end{eqnarray}
with ranks 3, 2, and 1 lead to the contradiction
$1\stackrel{!}{=}3+2-2$.  \bigskip

{\sc Proof:} As both operators $\bbox A _1$ and $\bbox A _2$ are
Hermitian, they have diagonal representations with respect to their
eigenvectors $\bbox A _1 |\lambda_{1,i}\rangle =
\lambda_{1,i}\,|\lambda_{1,i}\rangle$ and $\bbox A _2 |\lambda
_{2,j}\rangle = \lambda _{2,j}\,|\lambda _{2,j}\rangle$. Without loss
of generality we choose the normalized eigenvectors
$\{|\lambda_{1,i}\rangle\}$ as the basis of $\cal H$.

Clearly, $\cal H$ can be decomposed in the following two ways into
orthogonal subspaces
\begin{equation}
  \label{eq:decompose_space_a}
  {\cal H}=\ker(\bbox A _1) \oplus \text{Range}(\bbox A _1 )
\end{equation}
and also
\begin{equation}
  \label{eq:decompose_space_b}
  {\cal H}=\ker(\bbox A _2) \oplus \text{Range}(\bbox A _2 )\;.
\end{equation}
In a similar fashion we can partition the spectrum of $\bbox A _1$,
and hence $\cal H$'s basis, into all eigenvectors that form a basis of
$\text{Range}(\bbox A _1)$ and those that generate $\ker(\bbox A _1)$.
Since $\cal H$ is a complete linear space and $\bbox A _1, \bbox A _2,
\bbox A$ are linear operators, it is sufficient to consider the basis
states only. For any such state $|\lambda_{1,i}\rangle$ we find
\begin{eqnarray}
  \label{eq:basis_state}
  \langle\lambda_{1,i}|\bbox A|\lambda_{1,i}\rangle &=&
  \mu\langle\lambda_{1,i}|\bbox A _1|\lambda_{1,i}\rangle +
  \nu\langle\lambda_{1,i}|\bbox A _2|{\lambda}_{1,i}\rangle 
  \nonumber\\
  &=&  \mu\lambda_{1,i} + \nu{\Lambda}_i\;,
\end{eqnarray}
where we define the mean $ \Lambda_i=\langle\lambda_{1,i}|\bbox A
_2|{\lambda}_{1,i}\rangle = \sum_j\left| \langle\lambda_{2,j} |
  {\lambda}_{1,i}\rangle\right|^2{}\, {\lambda}_{2,j}\geq 0$. This
quantity is always positive (or zero) by virtue of $\bbox A _2$ being
positive semidefinite.

In accordance with the decomposition in
Eqs.(\ref{eq:decompose_space_a}) and (\ref{eq:decompose_space_b}) four
different cases are to be distinguished:
\begin{equation}
  \label{eq:fall_differentiation}
  \begin{array}{rl}
    |\lambda_{1,i}\rangle\in\text{Range}(\bbox A _1):\; 
    &\left\{\begin{array}{l@{\quad\Rightarrow\quad}l}
      |\lambda_{1,i}\rangle\notin\ker(\bbox A _2)
      & \langle\lambda_{1,i}|\bbox A |\lambda_{1,i}\rangle =  
      \mu\lambda_{1,i}+ \nu\Lambda_i >0
      \\
      |\lambda_{1,i}\rangle\in\ker(\bbox A _2)
      & \langle\lambda_{1,i}|\bbox A |\lambda_{1,i}\rangle =
      \mu\lambda_{1,i}+ 0  >0
    \end{array}\right.\\
    |\lambda_{1,i}\rangle\in\ker(\bbox A _1):\; 
    &\left\{
    \begin{array}{l@{\quad\Rightarrow\quad}lr}
      |\lambda_{1,i}\rangle\notin\ker(\bbox A _2)
      & \langle\lambda_{1,i}|\bbox A |\lambda_{1,i}\rangle =
      0 + \nu\Lambda_i >0
      \\
      |\lambda_{1,i}\rangle\in\ker(\bbox A _2)
      & \langle\lambda_{1,i}|\bbox A |\lambda_{1,i}\rangle = 0 + 0
    \end{array}\right.
  \end{array}
\end{equation}
Therefore only (basis) vectors that lie in {\em both} nullspaces will
belong to the nullspace of $\bbox A$, which proves the first part of
the Lemma. The second part follows from the linear algebraic dimension
relation
\begin{eqnarray}
  \label{eq:dimsion_theorem}
  \lefteqn{\dim\left(\,\text{Range}(\bbox A _1)
    +\text{Range}(\bbox A _2)\,\right)}\nonumber\\
  &&\qquad=\text{rank}(\bbox A _1) + \text{rank}(\bbox A _2)
  -\dim\left(\,\text{Range}(\bbox A _1)\cap\text{Range}(\bbox A _2)\,
  \right)\;,
\end{eqnarray}
where ``+'' on the lefthand side denotes all linear combinations of
the vectors in both ranges.

Now, any vector that lies either in $\text{Range}(\bbox A _1)$ or in
$\text{Range}(\bbox A _2)$ will, with an argument similar to
Eq.(\ref{eq:fall_differentiation}), always be in $\text{Range}(\bbox
A)$. We are thus allowed to replace
\begin{equation}
  \label{eq:rank_dim}
  \text{rank}(\mu\bbox A _1 
  + \nu\bbox A  _2 )=  
  \dim\left(\,\text{Range}(\bbox A _1)
    +\text{Range}(\bbox A _2)\,\right)
  \;,
\end{equation}
which completes our proof.

The values of $\mu,\nu>0$ are arbitrary, although often physical
constraints may suggest that some specific values may be better than
others (see the discussion in Section~\ref{sec:optimal_rec}).

We note that the lemma's first part could have been proved without
using a basis. The decomposition Eq.(\ref{eq:decompose_space_a}) and
the differentiation of Eq.(\ref{eq:fall_differentiation}) into
$|\phi\rangle\in \ker(\bbox A _1)$ or $|\phi\rangle\notin \ker(\bbox A
_1)$ for any $|\phi\rangle\in{\cal H}$ suffices. However, the second
part of the lemma requires the basis vectors.

\begin{table}[p]
  \caption{Characteristics of the initial wavepackets}
  \begin{center}
    \begin{tabular}{ldcdd}
      Configuration index & $x_0$ & $\langle\psi_0|H|\psi_0\rangle$ &
      \multicolumn{2}{c}{classical turning points}\\
      &&&left &right\\
      \tableline
      H& 1.75&  0.081& -2.1563& 2.1534 \\
      R& 0.9977&0.055& -2.0013& 1.9978 \\
      T& 0.0052&0.061& -2.0403& 2.0370 \\
      L&-1.002& 0.054& -1.9996& 1.9961 
    \end{tabular}
  \end{center}
  The configuration indices H, R, T, and K corresponding to the
  locations of wavepacket initial positions are shown in
  Fig.~\ref{fig:malonaldehyd}. All wavepackets start with equal
  width $\sigma=0.05$ and are initially at rest centered at the
  respective starting position $x_0$. The average energy of each
  packet as well as the corresponding turning points of an
  equivalent classical particle of the same energy are given.
  \label{tab:energy}
\end{table}

\begin{table}[p]
  \caption{Inversion regularization information}
  \begin{center}
    \begin{tabular}{lddddd}
      Configuration & $\alpha_1$ &$x_a$&$x_b$ & $|\Delta u|\times10^{-3}$ &
      $|\Delta s|\times10^{-3}$\\
      \tableline
      H$_1$    &3.3$\times10^{-5}$  &-4.0&4.0 &384.58 & 0.03
      \\
      H        &1.0                 &-2.0&2.0 &11.52  & 23.46
      \\
      R        &0.033       &-1.5&1.5 &7.16   &1.06
      \\
      T        &0.007      &-1.5&1.5 &9.02  &0.05
      \\
      L        &0.033     &-1.5&1.5 &6.53   &1.07
      \\[\medskipamount]
      $\Sigma$ &100.0   &-1.5&1.5 &9.53  &111.63
      \\
      LTRH     &0.333      &-1.5&1.5 &3.83   &12.42
      \\
      LTR      &0.01    &-1.5&1.5 &2.78   &0.70
      \\
      LT       &0.01      &-1.5&1.5 &3.10  &0.49
    \end{tabular}
  \end{center}
  In this numerical case study the optimal regularization parameter
  value $\alpha_1$ was identified by scanning its effect on the
  solution defect $|\Delta u|$. The inversion domains are
  $x_a\leqslant x\leqslant x_b$. The system defect is $|\Delta s|$.
  The first five rows apply to the individual PES reconstructions
  shown in Fig.~\ref{fig:individual_reconstruction}, and the last
  four rows refer to measurement combinations shown in
  Fig.~\ref{fig:combi_reconstruction}. See the text for details.
  \label{tab:scan}
\end{table}

\cleardoublepage

\begin{figure}[p]
  \begin{center}
    \includegraphics[width=8cm]{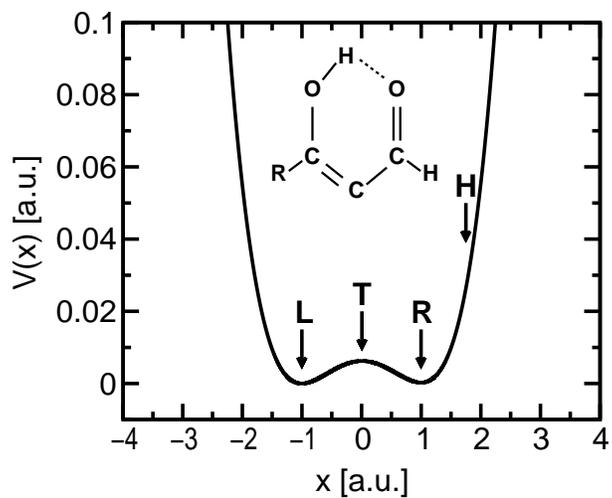}
  \end{center}
  \caption{The substituted malonaldehyde model system with its
    corresponding one dimensional potential energy function as given
    in Eq.(\ref{eq:double_well}). L, T, R, H indicate the different
    wavepacket initial positions utilized for the simulated
    inversions.}
  \label{fig:malonaldehyd}
\end{figure}

\begin{figure}[p]
  \begin{center}
    {\includegraphics[width=8cm]{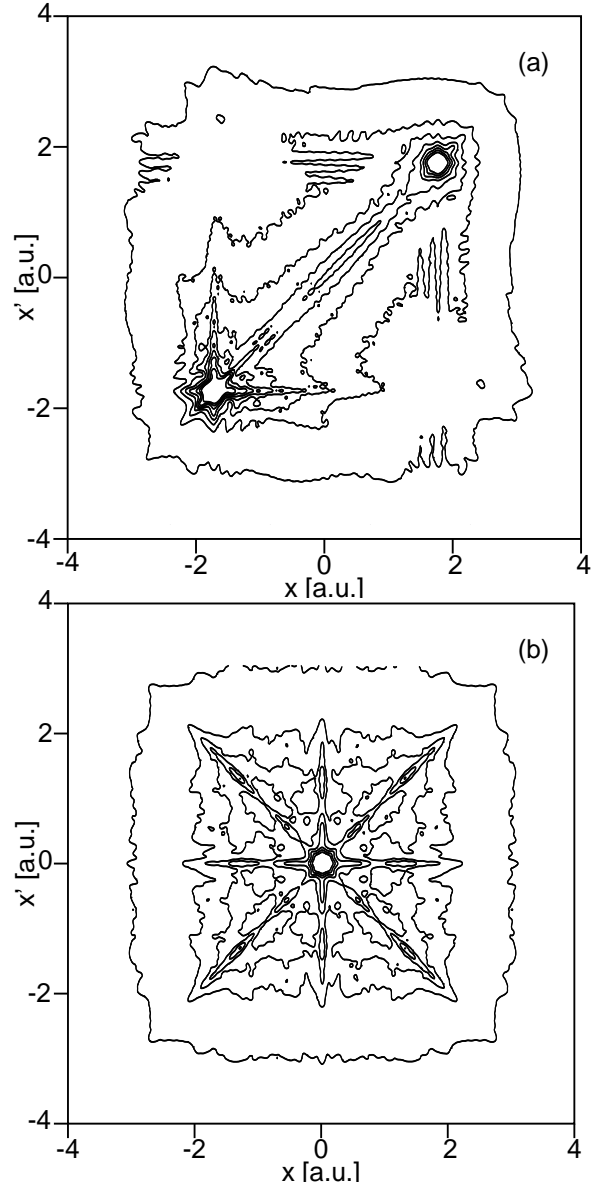}}
  \end{center}
  \caption{Contour plots of the kernel matrices $\bbox A$. (a)
    configuration H and (b) configuration T. The numerical values for
    the matrix entries range from $\thicksim10^3$ on the diagonal to
    $\thicksim10^{-8}$ on the boundaries. The contour levels
    correspond to: 1 (outer line), 31, 61, \ldots, 211.}
  \label{fig:kernel}
\end{figure}

\begin{figure}[p]
  \begin{center}
    {\includegraphics[width=8cm]{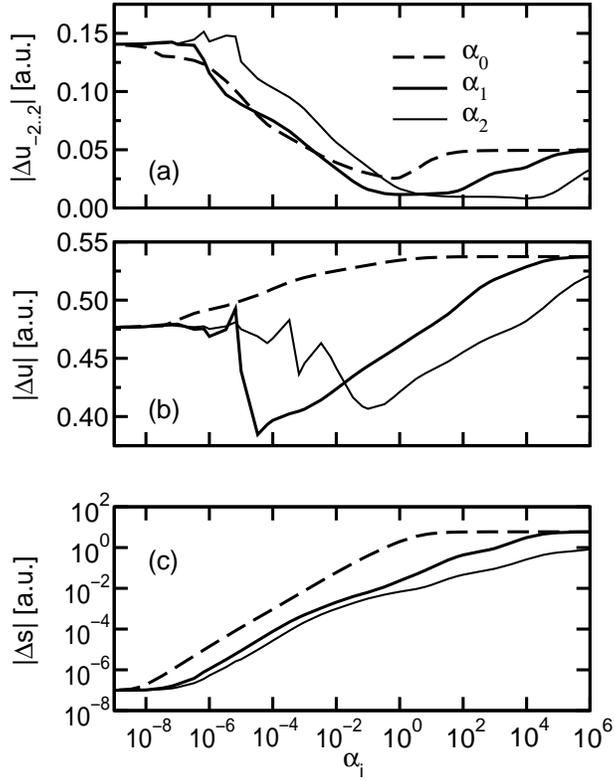}}
  \end{center}
  \caption{$\alpha_i$ parameter scans performed with configuration H. 
    Panels (a) and (b) display the solution defect $|\Delta u|$ with
    respect to two different inversion ranges: $-2\leqslant x\leqslant
    2$ and $-4\leqslant x\leqslant 4$, respectively. Panel (c) shows
    the system defect $|\Delta s|$ for the entire domain $-4\leqslant
    x\leqslant 4$.}
  \label{fig:scans}
\end{figure}

\begin{figure}[p]
  \begin{center}
    {\includegraphics[width=8cm]{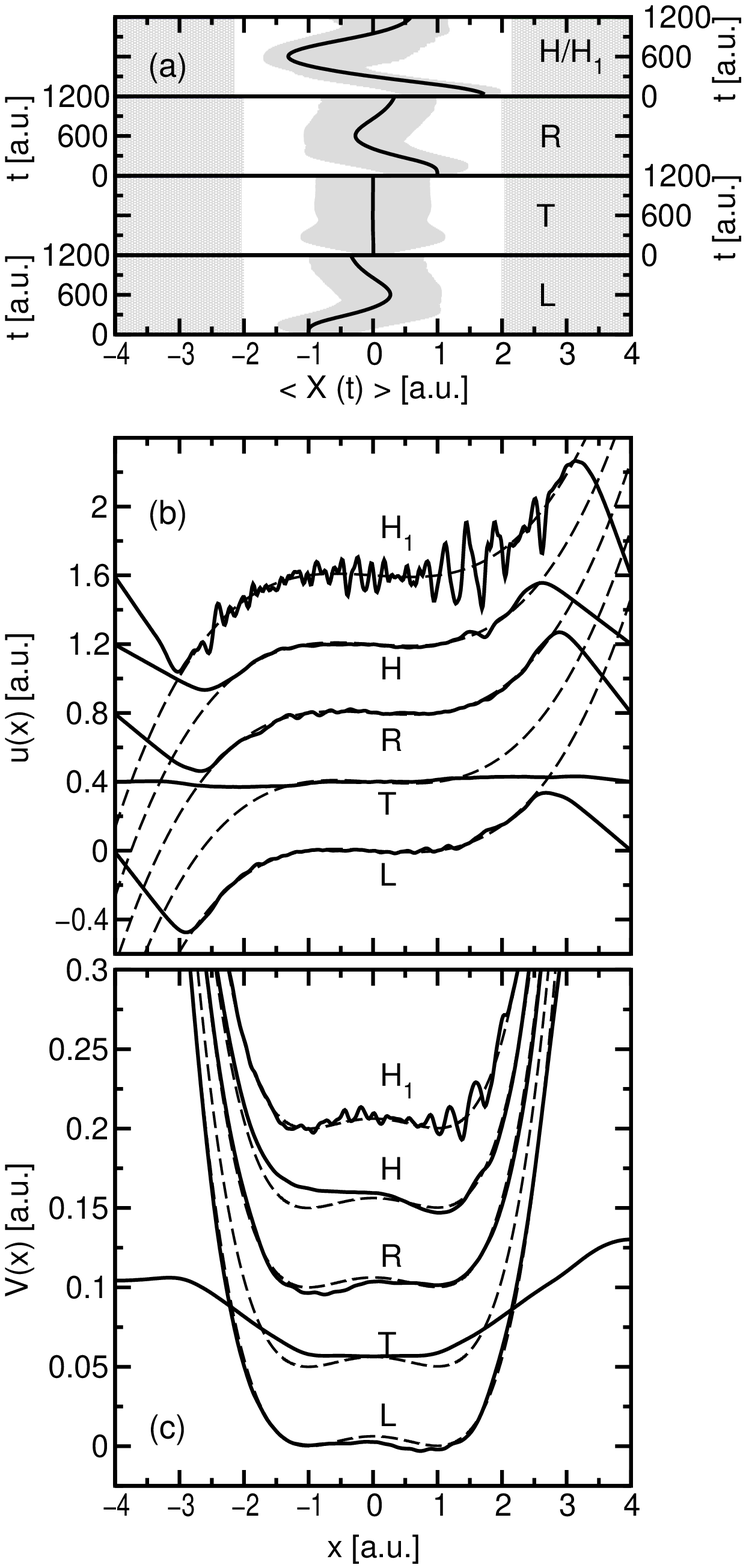}}
  \end{center}
  \caption[*]{}
  \label{fig:individual_reconstruction}
\end{figure}
\clearpage
FIG.~\ref{fig:individual_reconstruction}.  Extractions of the
potential under the conditions given in table~\ref{tab:scan}. (a) the
time evolution of the position average $\langle x(t)\rangle$
accompanied by the left- and righthand variance (i.e., shaded regions
bounded by Eqs.(\ref{eq:left_variance}) and~(\ref{eq:right_variance}))
to indicate the regions predominantly covered by the probability
densities. The grey domains on the extreme left and right mark
classically forbidden areas (cf.  table~\ref{tab:energy}).(b) the
reconstructed $u(x)$ and the corresponding potential $V(x)$ in (c)
with a suitably chosen additive constant. For comparison the exact
solutions are included as dashed lines. The individual curves have
been offset for graphical reasons and the detailed presentation of
$V(x)$ is restricted to $|x|\lesssim 2.5$ since the boundary regions
will not be extracted correctly due to lack of data sampling there.

\begin{figure}[p]
  \begin{center}
    {\includegraphics[width=7cm]{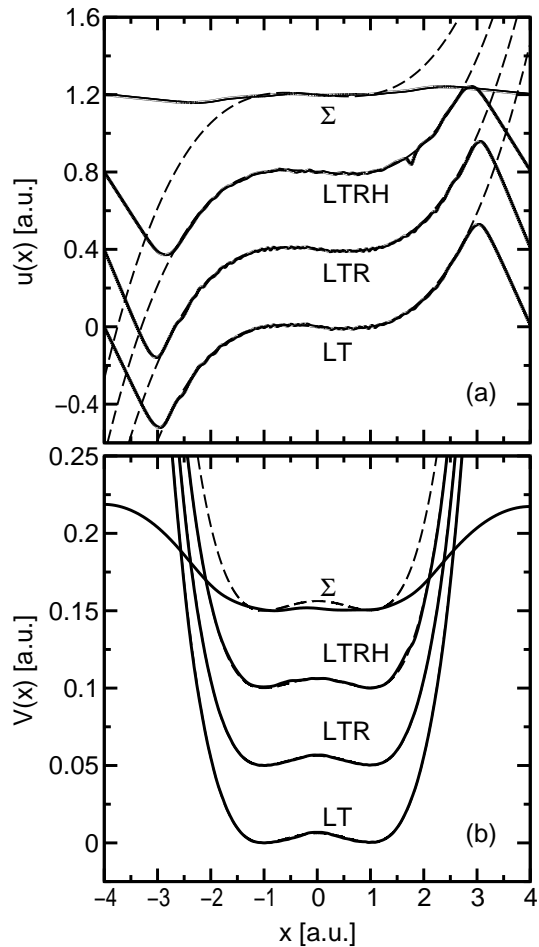}}
  \end{center}
  \caption{Extraction of the PES for optimally combined (LT, LTR, and
    LTRH) as well as $\rho$-combined data $(\Sigma)$. See the text and
    table~\ref{tab:scan} for details. The curves for the derivative
    $u(x)$ in (a) and the PES in (b) have been offset for graphical
    clarity and exact solutions (dashed lines) added for comparison.
    For optimal combinations of the data the original and
    reconstructed PES are almost indistinguishable.}
  \label{fig:combi_reconstruction}
\end{figure}



\begin{thebibliography}{10}

\bibitem{LevineBernstein}
R.~D. Levine and R.~B. Bernstein, {\em Molecular Reaction Dynamics} (Oxford
  Univ. Press, New York, N.Y., 1974).

\bibitem{BornOppenheimer}
M. Born and J.~R. Oppenheimer, Annalen der Physik {\bf 84},  457  (1927).

\bibitem{Hartree}
D.~R. Hartree, Proceedings of the Cambridge Philosophical Society {\bf 24},  89
   (1928).

\bibitem{Fock1}
V.~A. Fock, Zeitschrift f{\"u}r {P}hysik {\bf 61},  126  (1930).

\bibitem{Fock2}
V.~A. Fock, Zeitschrift f{\"u}r {P}hysik {\bf 62},  795  (1930).

\bibitem{SzaboOstlund}
A. Szabo and N.~S. Ostlund, {\em Modern Quantum Chemistry: Introduction to
  Advanced Electronic Structure Theory} (Macmillan Publishing, New York, 1982).

\bibitem{Morse_PhysRev1929}
P.~M. Morse,  Physical Review \textbf{34}, 57  (1929).

\bibitem{Gerber_PhysRevLett1978}
R.~B. Gerber, M. Shapiro, U. Buck, and J. Schleusener, Physical Review Letters
  {\bf 41},  236  (1978).

\bibitem{Lowe_SIAM1992}
B. Lowe, M. Pilant, and W. Rundell, SIAM Journal on Mathematical Analysis {\bf
  23},  482  (1992).

\bibitem{Ho_JPhysChem1993}
T.-S. Ho and H. Rabitz, Journal of Physical Chemistry {\bf 97},  13447  (1993).

\bibitem{GerambInversionTheory}
{\em Quantum Inversion Theory and Applications}, edited by H.~V. von Geramb
  (Springer, New York, N.Y., 1994).

\bibitem{Fabiano_IMA1995}
R. Fabiano, R. Knobel, and B. Lowe, IMA Journal or Numerical Analysis {\bf 15},
   75  (1995).

\bibitem{Zhang_JChemPhys1995}
D.~H. Zhang and J.~C. Light, Journal of Chemical Physics {\bf 103},  9713
  (1995).

\bibitem{Ho_JChemPhys1996}
T. Ho, H. Rabitz, S. Choi, and M. Lester, Journal of Chemical Physics {\bf
  104},  1187  (1996).

\bibitem{Zewail_JPhysChem1993}
A. Zewail, Journal of Physical Chemistry {\bf 97},  12427  (1993).

\bibitem{Shapiro_JPhysChem1996}
M. Shapiro, Journal of Physical Chemistry {\bf 100},  7859  (1996).

\bibitem{Geiser_JChemPhys1998}
J.~D. Geiser and P.~M. Weber, Journal of Chemical Physics {\bf 108},  8004
  (1998).

\bibitem{SchroedingerGleichung}
E. Schr{\"o}dinger, Annalen der Physik {\bf 79},  361  (1926).

\bibitem{Zhu_JChemPhys1999Jul}
W. Zhu and H. Rabitz, Journal of Chemical Physics {\bf 111},  472  (1999).

\bibitem{Zhu_JPhysChemA1999}
W. Zhu and H. Rabitz, Journal of Physical Chemistry A {\bf 103},  10187
  (1999).

\bibitem{Vager_Science1989}
Z. Vager, R. Naaman, and E.~P. Kanter, Science {\bf 244},  426  (1989).

\bibitem{Kwon_PhysRevLett1996}
K. Kwon and A. Moscowitz, Physical Review Letters {\bf 77},  1238  (1996).

\bibitem{Assion_PhysRevA1996}
A. Assion {\it et~al.}, Physical Review A {\bf 54},  R4605  (1996).

\bibitem{Williamson_Nature1997}
J.~C. Williamson {\it et~al.}, Nature {\bf 386},  159  (1997).

\bibitem{Krause_PhysRevLett1997}
J.~L. Krause, K.~J. Schafer, M. Ben-Nun, and K.~R. Wilson, Physical Review
  Letters {\bf 79},  4978  (1997).

\bibitem{Jones_PhysRevA1998}
R.~R. Jones, Physical Review A {\bf 57},  446  (1998).

\bibitem{EhrenfestTheorem}
P. Ehrenfest, Zeitschrift f{\"u}r Physik {\bf 45},  455  (1927).

\bibitem{Zhu_JPhysChemB2000}
W. Zhu and H. Rabitz, Journal of Physical Chemistry B {\bf 104},  10863
  (2000).

\bibitem{Louis_GAMM1990}
A.~K. Louis, GAMM-Mitteilungen {\bf 1},  5  (1990).

\bibitem{NumericalRecipesC}
W.~H. Press, S.~A. Teukolsky, W.~T. Vetterling, and B.~P. Flannery, {\em
  Numerical Recipes in C: The Art of Scientific Computing}, 2nd ed. (Cambridge
  University Press, ADDRESS, 1994).

\bibitem{Doslic_JPhysChem1998}
N. Do{\v{s}}li{\'c}, O. K{\"u}hn, J. Manz, and K. Sundermann, Journal of
  Physical Chemistry A {\bf 102},  9645  (1998).

\bibitem{Feit_JApplOpt1978}
M.~D. Feit and J.~A. Fleck, Applied Optics {\bf 17},  3990  (1978).

\bibitem{Feit_JCompPhys1982}
M.~D. Feit, J.~A. Fleck, and A. Steiger, Journal of Computational Physics {\bf
  47},  412  (1982).

\bibitem{TikhonovJohn}
A.~N. Tikhonov and F. John,  in {\em Solutions of Ill-Posed Problems}, {\em
  Scripta Series in Mathematics}, edited by V. Arsenin (Winston, Washington,
  D.C., 1977).
  
\bibitem{foot} For $b(x)$, a related issue pointed out
  in~\cite{Zhu_JChemPhys1999Jul} is the stability of $b(x)$ in view of
  the need to take the second time derivative of the probability
  density.  An approach based on partial integration over time has
  been proposed calling for a first order time derivative only.
  However, a check of the inversion performance based on partial
  integration produced unsatisfactory results. It will always be
  extremely difficult to reliably compute the terms
  \begin{displaymath}
    \left.
      -\frac{m}{T}\rho(x,t)\,\frac{\D}{\D t} \langle x(t) \rangle 
    \right|_0^T
  \end{displaymath}
  at only a few $\rho(x,t)$ snapshots in time. One inevitably needs to
  work with one-sided derivatives at $t=0$ and $T$, which
  significantly diminishes the accuracy.

\end{thebibliography}
\end{document}